\documentstyle[12pt]{article}

\newcommand{\be}{\begin{equation}}
\newcommand{\ee}{\end{equation}}

\newcommand{\ba}{\begin{eqnarray}}
\newcommand{\ea}{\end{eqnarray}}

\newcommand{\la}{\lambda}
\newcommand{\al}{\alpha}

\newcommand{\r}{\rho}

\begin{document}

\hsize36truepc\vsize51truepc
\hoffset=-.4truein\voffset=-0.5truein
\setlength{\textheight}{8.5 in}

\begin{titlepage}
\begin{center}
\hfill LPTENS-99/49\\
\hfill November, 1999\\

\vskip 0.6 in
{\large  Logarithmic moments of characteristic polynomials of random
matrices}
\vskip .6 in
       {\bf Edouard Br\'ezin \footnote{$  ${\it
 Laboratoire de Physique Th\'eorique de l'\'Ecole Normale
Sup\'erieure, Unit\'e Mixte de Recherche 8549 du Centre National de la
Recherche
Scientifique et de l'\'Ecole Normale Sup\'erieure,
24 rue Lhomond, 75231 Paris Cedex 05, France.\\ {\bf
brezin@physique.ens.fr}}}}
            {\it{and}}\hskip 0.3cm
       {\bf {Shinobu Hikami}}  \footnote{ $  ${\it{ Department of
Basic Sciences, University of Tokyo, Meguro-ku, Komaba 3-8-1, Tokyo
153,
Japan.\\ {\bf hikami@rishon.c.u-tokyo.ac.jp}}}}\\

\vskip 0.6 in
{\it{Dedicated for his birthday to our very distinguished colleague and
dear friend Joel Lebowitz, the scientist and the
untamable militant for  human rights}}
 \vskip 0.5 cm
{\bf Abstract}
\end{center}
\vskip 14.5pt

{\leftskip=0.5truein\rightskip=0.5truein\noindent
{\small

In a recent article we have discussed the connections between averages of
powers of
Riemann's $\zeta$-function on the critical line, and averages of
characteristic
polynomials of random matrices. The result for random matrices was shown to
be universal, i.e.
independent of the specific probability distribution, and the  results were
derived for
arbitrary moments. This allows one to extend the previous results to
logarithmic moments,
for which we derive the explicit universal expressions in random matrix
theory. We then compare these
results to various results and conjectures for $\zeta$-functions, and the
correspondence
is again striking.

}
\par}
\end{titlepage}
\setlength{\baselineskip}{1.5\baselineskip}
\section{  Correlation functions of characteristic polynomials}
We first briefly review the result of a previous paper \cite{BH1},
in which we have investigated the average of a product of
 characteristic polynomials of a random matrix.
Let $X$ be an $M \times M$ random Hermitian matrix.
The correlation function of $2K$ distinct characteristic
polynomials is defined as
\be F_{2K}({\la}_1,\cdots,{\la}_{2K}) = \langle
\prod_{{\al}= 1}^{2K}\det({\la}_{{\al}}-X)\rangle.\ee
The average is taken with  the  normalized
probability distribution
\be \label{weight} P(X) =\frac{1}{Z} \exp -N {\rm Tr} V(X), \ee
where $V(X)$ is a polynomial in $X$.
The simplest case consists of a  Gaussian distribution, and the result is
easily
worked out by the orthogonal polynomial method,
\be \label{GAUSS} P(X) = \frac{1}{Z_M} \exp - \frac{N}{2} \rm Tr X^2
,\ee
with
\be M= N-K.\ee
Defining the middle-point
\be \la = \frac {1}{2K}\sum_{\al= 1}^{2K} \la_\al,\ee
and the density of eigenvalues at this point
\be \r(\la) = \frac{1}{2\pi} \sqrt{4-\la^2}, \ee
we introduced the scaling variables
\be \label{scaling} x_a = 2\pi  N \r(\la) (\la_a-\la), {\rm{with}}
\sum_{a= 1}^{2K} x_a= 0, \ee
and consider the large N limit, finite $x_a$'s limit. In this limit we have
shown \cite{BH1} that,
\ba \label {final}&&
\exp-{(\frac{N}{2}\sum_{l=1}^{2K}V(\la_l))}F_{2K}(\la_1,\cdots,\la_{2K}
) =
\nonumber \\&&(2\pi N\r(\la))^{K^2}
\frac{\exp(- NK)}{K!} \oint \prod_1^K \frac
{du_{\al}}{2\pi}\exp{-i(\sum_{\al=1}^K u_{\al})}\
\frac{\Delta^2(u_1,\cdots,u_K)}
{\prod_{\al=1}^K\prod_{l=1}^{2K} (u_{\al}-x_l)}. \ea
in which the contours enclose all the $x_l$'s. It has also been shown that
the result
is universal in this scaling limit, i.e. independent of the specific
polynomial $V$ which defines the probability distribution.
For K = 1, we have
\be \label {finalx}
\exp\{-{\frac{N}{2}(V(\la_1)+V(\la_2))}\}F_{2}(\la_1,\la_{2}) = e^{-N}
(2\pi N\r(\la)) \frac{\sin x}{x}\ee
with $x= \pi N \r(\la) (\la_1-\la_2)$,
the familiar sine-kernel.

When all the $\la_i$'s are equal, we obtain the $2K$-th moment of the
characteristic polynomial :
\ba &&\label{final2} \exp-{(NKV(\la))}F_{2K}(\la,\cdots,\la) = \nonumber\\
 && (2\pi
N\r(\la))^{K^2}
\frac{\exp(-NK)}{K!} \oint \prod_1^K \frac
{du_{\al}}{2\pi}\exp{-i(\sum_{\al=1}^K u_{\al})}\
\frac{\Delta^2(u_1,\cdots,u_K)}
{\prod_{\al=1}^K u_{\al}^{2K}}. \ea
This contour integration reduces to a simple determinant
and one finds,
\be \label{moment} \exp-{(NKV(\la))}F_{2K}(\la,\cdots,\la) = (2\pi
N\r(\la))^{K^2}
e^{-NK}\prod_0^{K-1}\frac{l!}{(K+l)!} . \ee

Let us denote the last factor by
\be\label{gammaK}
\gamma_K = \prod_0^{K-1}\frac{l!}{(K+l)!}.
\ee
 In our previous paper \cite{BH1}, we  have compared (\ref{moment}) with the
average of the $2K$-th moment of the $\zeta$-function
\cite{Keating,Conrey},
for which it has been conjectured  that
\be\label{conj}
\frac{1}{T} \int_0^T dt | \zeta( \frac{1}{2} + i t) |^{2 K}
\simeq \gamma_K a_K  (\log T)^{K^2}
\ee
where $a_K$ is a number theoretic coefficient given by the
product of the prime $p$,
\be
a_K = \prod_{p} [ ( 1 - \frac{1}{p})^{K^2}
\sum_{m=0}^{\infty} ( \frac{K(K+1)\cdots (K+m-1)}{m!})^2 p^{-m}],
\ee
and $\gamma_K$ is the same as in (\ref{gammaK}). Since we have with
(\ref{moment}) an expression valid for all $K$'s, we shall extend this
comparison
beween the moments of characteristic polynomials and that of zeta-functions
to non-integer K,
as will be explained in the next section.

\section{ Non-integer power moment }

Since the result (\ref{moment}) is valid for any $K$ we may now  consider
the analytic continuation to non-integer $K$.
This requires to continue the coefficient  $\gamma_K$ in (\ref{gammaK})
to non-integer values.
This will be needed  for obtaining the logarithmic moment of the
characteristic polynomials.
The non-integer power moments are also interesting by themselves, since
there exists  equivalent studies of  fractional power moments of the
Riemann $\zeta$-function \cite{Heath-Brown}.

The  factor  $\gamma_K = \prod l!/(K + l)!$ may be
expressed through an integral representation,
\be\label{int}
\log \gamma_K = - \int_0^{\infty} dt \frac{e^{-t}}{t} [ K^2 - \frac{(1 -
e^{-Kt})^2}
{(1 - e^{-t})^2}]
\ee
which is easily checked by expanding the integrand in powers of $e^{-t}$.
It is more conveniently handled if we take it as
\be\label{int2}
\log \gamma_K = \lim_{\alpha \rightarrow 1} - \int_0^{\infty}
dt \frac{e^{-t}}{t^\alpha} [ K^2 - \frac{(1 - e^{-Kt})^2}
{(1 - e^{-t})^2}],
\ee
since it may then be splitted into two parts for $\alpha<1$.

Expanding the integrand in the power of $e^{-t}$, and integrating over $t$,
we obtain
\ba
\log \gamma_K &=& - \Gamma(1 - \al)[K^2 - \sum_{n=0}^{\infty} (n + 1)^{
\al}
+ 2 \sum_{n=0}^{\infty} ( n + K + 1)^{\al - 1}(n + 1)\nonumber\\
&& - \sum_{n=0}^{\infty}( n + 2K + 1)^{\al - 1}(n + 1)]\nonumber\\
&=&- \Gamma(1 - \al) [ K^2 - \zeta(-\al) + 2 \zeta(-\al,2K + 1) - 2 K
\zeta(1 - \al,K + 1)
\nonumber\\
&& - \zeta(-\al,2 K + 1) + 2 K \zeta(1 - \al, 2 K + 1)]
\ea
in which the limit $\al \rightarrow 1$ is meant.
The generalized zeta-function $\zeta(z,a)$ is
given by $  \sum_{n=0}^{\infty} ( a + n)^{- z}$, and $\zeta(0,a) =
\frac{1}{2} - a$ ;
it has the expansion
\ba
\zeta(z,a) &=& \frac{2 \Gamma( 1 - z)}{(2 \pi )^{1 - z}}[
\sin ( \frac{\pi z}{2})\sum_{n=1}^{\infty}
\frac{\cos ( 2 n \pi a)}{n^{1 - z}}
\nonumber\\
&+& \cos (\frac{\pi z}{2})\sum_{n=1}^{\infty} \frac{\sin ( 2 n \pi
a)}{n^{1 - z}}].
\ea

It is then easy to obtain various results for non-integer $K$. For instance
the $K=0$ limit is obtained by expanding in powers of $K$. The
term of  order
$K^2$
is
\ba
\log \gamma_K &\simeq& - K^2 [ \Gamma( 1 - \al) - \Gamma(3 -
\al)\sum_{n=0}^{\infty}
\frac{1}{( n + 1)^{2 - \al}}]\nonumber\\
&=& - K^2 [ \Gamma(1 - \al) - \Gamma(3 - \al) \zeta( 2 - \al)]\nonumber\\
&=& K^2 (1 + c) + O(K^3)
\ea
where $c$ is Euler's constant, $c = 0.5772..$.

It is also interesting to evaluate $\gamma_K$ for $K = \frac{1}{2}$, since
it is needed for computing the first moment of the characteristic
polynomial or of the Riemann
$\zeta$-function.
In this case, we have
\ba
\log \gamma_{\frac{1}{2}} &= &\lim_{\alpha\rightarrow  1} - \int_0^{\infty}
(\frac{1}{4} - \frac{1}{(1 +
e^{-t/2})^2})
\frac{e^{-t}}{t^\al} dt\nonumber\\
&=&\lim_{\alpha\rightarrow  1}- \frac{1}{4}\Gamma(1 - \al) - 2^{1 -
\al}\sum_{n=0}^{\infty}
(-1)^n [ (n + 2)^\al - (n + 2)^{\al - 1} ] \Gamma(1 - \al)
\nonumber\\
&=&\lim_{\alpha\rightarrow  1}
[ - \frac{1}{4} + 2^{1 - \al} ((2^{1 + \al} - 1)\zeta(-\al)
-(2^{\al} - 1)\zeta( 1 - \al)\Gamma(1 - \al)\nonumber\\
&=&\frac{1}{12}\log 2 + \frac{1}{2} \log \pi + 3 \zeta^{\prime} (-1)
\ea
since $\zeta(0) = -\frac{1}{2}$,
$\zeta(-1) = - \frac{1}{12}$, and $\sum_{n=0}^{\infty} (-1)^n
(n + 2)^{-s} = 1 + (2^{1 - s} - 1)\zeta(s)$.
This leads to  $\gamma_{\frac{1}{2}}$ is 1.1432....
In the literature on
Riemann $\zeta$-functions  concerning the moments (\ref{conj}), bounds have
been conjectured
for $0\le K \le 1$ \cite{Heath-Brown,Conrey-Farmer}, and they amount for
the equivalent of $\gamma_K$ to
\be\label{bound}
\frac{1}{\Gamma(K^2 + 1)} \le \gamma_K \le \frac{2}{\Gamma(K^2 + 2) (2 - K)}.
\ee
We find that our result of (\ref{int}) indeed satisfies this bound for $0
\le K \le
1$. For instance, the bounds require  $1.1033 \le \gamma_{\frac{1}{2}} \le
1.1768 $,
and we have found $\gamma_{\frac{1}{2}} = 1.1432$.  It is easy to verify this
bound by expanding  around $K=1$ and $K=0$, and
it does support the conjecture (\ref{bound}).

\section{ Moment of the logarithm}

We now consider the case for which  all the $\la_\al$'s are equal, and
expand the $2K$-th moment
in powers of $K$ :
\ba I &=& \label{eq1}F_{2K}(\la,\cdots,\la) e^{-N K V(\la) + N
K}
\nonumber\\
 &=& \langle
[\det({\la}-X) e^{-\frac{N}{2} V(\la) + \frac{N}{2}}]^{2K}\rangle
\nonumber\\
&=& \sum_{p=0}^{\infty} \frac{(2K)^p}{p!}\langle
[\log |\det (\la - X)| - \frac{N}{2}V(\la) + \frac{N}{2}]^p \rangle.
\ea
>From (\ref{moment}), we know the same $I$ as
\be\label{eq2}
I = \sum_{p^{\prime}=0}^{\infty} \frac{K^{2p^{\prime}}}{p^{\prime}!}
[\log (2 \pi N \rho(\la)]^{p^{\prime}}
\ee
provided we neglect, for the moment, the factor $\gamma_K$ and set it
equal to one.
We have thus made an  analytic continuation from integer $K$ to a
real variable $K$,
and we have
expanded in $K$.
We have seen in the previous section that, for $K$ small,
\be
 \gamma_K = 1 + K^2 (1 + c) + O(K^3)
\ee
where $c$ is Euler constant.
This  correction of order $K^2$ gives only a subleading term in (\ref{eq2})
compared to $\log ( 2 \pi N \rho(\la)$, and thus we were justified to
neglect it.

Comparing (\ref{eq1}) and (\ref{eq2}), we find
\be\label{logmoment1}
\langle
{\large [}\log | \det (\la - X)| - \frac{N}{2}V(\la) +
\frac{N}{2}{\large ]}^{2m}
\rangle
\simeq \frac{1}{2^{2m}} \frac{2m!}{m!}[ \log (2 \pi N \rho(\la)]^{m}
\ee

Since there is the expansion (\ref{eq2}) contains only even powers of $K$,
we have also,
\be\label{eq4}\label{eq3}
\langle
[\log | \det (\la - X)| - \frac{N}{2}V(\la) + \frac{N}{2}]^{2m +1}\rangle
= 0
\ee
It is instructive to verify this for m = 0 ; taking a derivative of
(\ref{eq4}) with respect to $\la$ one should verify that
\ba
\langle {\rm Tr}(\frac{1}{\la - X})\rangle &=& \int da
\frac{\rho(a)}{\la - a}\nonumber\\
&=&\frac{N}{2}V'(\la)
\ea
which is nothing but the saddle point equation of the large N limit which
determines $\rho(\la)$ \cite{BIPZ}.
The solution of this Riemann-Hilbert problem is expressed
through $G(z) = \frac{1}{N}<{\rm Tr} (\frac{1}{z - X})> =
(\frac{1}{2}V'(z) - P(z)\sqrt{Q(z)})$, in which $P$ and $Q$ are polynomials
fixed by the requirement that $G$ falls like $1/z$ at infinity. Then  Re
$G(x+i\epsilon) = V'(x)$,
Im $G(x+i\epsilon) = -\pi\rho(x)$  in the large N limit. The r.h.s. of the
above equation is
indeed $N {\rm Re} G(\la) = \frac{N}{2} V'(\la)$.

Our argument may be applied to the moment of the $\zeta$-function
as well.
The average of
$2K$ moment of $\zeta$ function has been conjectured \cite{Keating,Conrey}
as
\be
\frac{1}{T} \int_0^T dt |\zeta(\frac{1}{2} + i t)|^{2K} \simeq \gamma_K
a_K (\log T)^{K^2}
\ee
Expanding the above equation in powers of $K$, we obtain along the same
lines
\be\label{eq5}
\frac{1}{T} \int_0^T [ \log |\zeta(\frac{1}{2} + i t)|]^{2m} \simeq
\frac{1}{2^{2m}}
\frac{2m!}{m!}(\log \log T)^m
\ee
provided $a_K$ goes to one in the limit $K\rightarrow 0$.
Selberg \cite{Selberg} has derived
\be\label{eq6}
\frac{1}{T} \int_0^T [ {\rm Im}\log (\zeta(\frac{1}{2} + i t))]^{2m}
\sim \frac{1}{2^{2m}}
\frac{2m!}{m!}(\log \log T)^m
\ee
where ${\rm Im}\log \zeta(x) = {\rm arg}\zeta(x)$.

Our result  (\ref{logmoment1}) may  be compared to (\ref{eq5}) or
(\ref{eq6}), in which
the density of state $2 \pi N \rho(\la)$ is replaced by $\log T$.
The characteristic polynomial $\det(\la - X)$ has zeros on the real axis
 in the complex plane of $\la$.
Hence this function  corresponds
to Riemann's $\xi (\frac{1}{2} + i \la) =
A(\la) Z(\la)$, which has zeros on the real $\la$ line, where
\be
A(\la) = \pi^{-\frac{1}{4}}e^{{\rm Re}\log \Gamma(\frac{1}{4} +
\frac{i \la}{2})}
( - \frac{\la^2}{2} - \frac{1}{4}),
\ee
\be
Z(\la) = e^{i \theta} \zeta(\frac{1}{2} + i \la).
\ee
 The analogy between $|\det(\la - X)|$ and $|\xi(\frac{1}{2} + i \la)| =
 |A(\la)| |\zeta(\frac{1}{2} + i \la)|$ leads to
 $\log|\zeta(\frac{1}{2} + i \la)| \sim \log|\det(\la - x)| - \log|A(\la)|$,
 and it may thus correspond to $\log|\det(\la - X)| - \frac{NV}{2} +
\frac{N}{2}$.

 From those log-moment results, we find that the distribution function
is a normal Gaussian
distribution. For the $\zeta$ function, or more generally for the $L$
functions,
it is known that
$\log |L(\frac{1}{2} + i t)|$ is ditributed like a random variable, when
$t$ is large,
with a Gaussian density \cite{Selberg,Bombieri,Laurincikas}.
 The coefficient of (\ref{logmoment1}),
 $2m!/(2^{2m} m!)$ is equal to  $ \frac{1}{\sqrt{\pi}}
\Gamma(m + \frac{1}{2})$, which is identical to the coefficient of
the following Gaussian integral,
\be
\int_{-\infty}^{\infty} x^{2m} e^{- a x^2} dx = \frac{1}{\sqrt{\pi}}
\Gamma( m + \frac{1}{2}) a^{-m - \frac{1}{2}}.
\ee
Hence we find that our moment (\ref{logmoment1}) does follow a normal
distribution.

We have considered up to now the case of all the $\la_i$'s equal ;  in the
following we
shall consider
two different $\la_i$'s, and it will also appear  that
a  normal distribution
holds for the logarithmic moments.

\section{ Moments at two different points }

The  formula for $F_{2K}$ in (\ref{final}) provides
 also the correlation for  two different values of the "energies",
$\la_1$ and $\la_2$, if we set  $\la_a = \la_1 = \cdots = \la_K$ and
$\la_b = \la_{K+1} = \cdots = \la_{2K}$.
For instance, when $K= 2$, ($ l_1 =2, l_2 = 2$, and $l_1 + l_2 = 2K$), we
have $x_1 =
x_2 = - x_3 = - x_4 = x$.
Then, the contour integral in (\ref{final}) becomes
\be
\oint \frac{du_1 du_2}{(2 \pi )^2}
\frac{e^{-i (u_1 + u_2)} (u_1 - u_2)^2}{(u_1 - x)^2 (u_1 + x)^2 (u_2 -
x)^2 (u_2 + x)^2}
= \frac{1}{2x^2}( 1 - \frac{\sin^2x}{x^2})
\ee
where $x = \pi N \rho(\la) (\la_1 - \la_2)$, with $ \la =
\frac{1}{2}(\la_1 + \la_2)$.
Note that the contour integral formula of (\ref{final}) has been derived in
Dyson's short
distance limit \cite{Mehta,Dyson}.
We consider, within this Dyson limit, the large x case :
$x>>1$.
The leading term forlarge $x$, is easily obtained from
(\ref{final}).
The subleading terms have an oscillatory behavior, but we limit ourselves
for simplicity to the
leading term .
Then the leading behavior becomes
\be\label{l1l2}
 \langle [\det(\la_1 - X)]^{2 l_1} [\det(\la_2  - X)]^{2 l_2} \rangle
e^{- N l_1 V(\la_1) + l_1 N - N l_2 V(\la_2) + l_2 N}
\simeq \frac{1}{x^{\frac{1}{2}(\l_1 + \l_2)^2}} (2 \pi N \rho )^{(l_1 +
l_2)^2}
\ee
Expanding in powers  of $l_1$ and $l_2$, we obtain
from the l.h.s. of (\ref{l1l2}),
\ba
I &=& \sum_{p_1,p_2 = 0}^{\infty} \frac{(2 l_1)^{p_1}}{p_1!}
\frac{(2 l_2)^{p_2}}{p_2!} \langle [\log |\det(\la_1 - X)| -
\frac{N}{2} V(\la_1) + \frac{N}{2}]^{p_1}
[\log |\det (\la_2 - X)|\nonumber\\
& -& \frac{N}{2} V(\la_2) + \frac{N}{2}]^{p_2}
\rangle
\ea
For the r.h.s. of (\ref{l1l2}), we have
\be
\sum_{p=0}^{\infty}  (- \frac{1}{2})^p \frac{(l_1 + l_2)^{2p}}{p!} (\log
x)^p
=\sum_{p=0}^\infty \sum_{p_1,p_2, 2p = p_1 + p_2}
\frac{(2p)!}{p!} \frac{l_1^{p_1}}{p_1!} \frac{l_2^{p_2}}{p_2!}
[\log(
\frac{2\pi N \rho}{\sqrt{2 x}} )]^p
\ee
Hence, we obtain when $p_1 + p_2 $ is even integer,
\ba\label{p1p2}
&&\langle [\log |\det(\la_1 - X)| -
\frac{N}{2} V(\la_1) + \frac{N}{2}]^{p_1}
[\log |\det (\la_2 - X)| - \frac{N}{2} V(\la_2) + \frac{N}{2}]^{p_2}
\rangle\nonumber\\
&=& (\frac{1}{2})^{2 p} \frac{(2 p)!}{p!} [\log
(\frac{ 2 \pi N \rho }{\sqrt{2 x}})]^p
\ea
where $p = (p_1 + p_2)/2$.
When $p_1 + p_2$ is an odd integer, the correlation  vanishes since there
is no corresponding
term in the r.h.s..
The above result has been derived for  large $x$. In this limit,
we have obtained,
\ba\label{logmoment}
&&\langle [\log |\det(\la_1 - X) - \frac{N}{2}V(\la_1) - \log |\det(\la_2 -
X)|
+ \frac{N}{2}
V(\la_2)]^{2p}\rangle \nonumber\\
&\simeq&  \langle [\log |\det(\la_1 - X) - \log |\det(\la_2 - X)|]^{2p}
\rangle \nonumber\\
&\simeq& 2 (\frac{1}{2})^{2 p} \frac{(2p)!}{p!} (\log[2 \pi N \rho])^p
 - 2(\frac{1}{2})^{2 p} \frac{(2p)!}{p!} (\log[\frac{2 \pi N \rho}
{\sqrt{2x}}])^p
\ea
where we have expanded the binomial forms, and used
(\ref{p1p2}). The difference $V(\la_1) - V(\la_2)$ gives a subleading
term, and it has been neglected.

In the simple case, $p= 1$ for (\ref{logmoment}), we have a cross-term
with two logarithms. By taking the derivatives of this cross term, we
obtain
\be
\partial_{\la_1}\partial_{\la_2}
\langle \log|\det(\la_1 - X)| \log|\det(\la_2 - X)|\rangle
= \langle {\rm Tr} \frac{1}{\la_1 - X} {\rm Tr} \frac{1}{\la_2 - X}
\rangle
\ee
which is two point Green function. We have
\be
\langle {\rm Tr} \frac{1}{z_1 - X} {\rm Tr} \frac{1}{z_2 - X}
\rangle
= N^2 G_{2c}(z_1,z_2) + N^2 G(z_1) G(z_2)
\ee
where the connected two-point Green function $G_{2c}(z_1,z_2)$ has been foun=
d
in \cite{BZ,BHZ}. There it has been shown that
\ba\label{G2}
N^2 G_{2c}(z_1,z_2) &=& - \partial_{z_1}\partial_{z_2}
\log [ 1 - G(z_1) G(z_2)]\nonumber\\
&=&\partial_{z_1}\partial_{z_2} \log [ \frac{u(z_1) - u(z_2)}
{z_1 - z_2}]\nonumber\\
&=& \frac{1}{2(z_1 - z_2)^2} ( \frac{z_1 z_2 - 4}{[(z_1^2 - 4)(z_2^2 -
4)]^{1/2}
} - 1)
\ea
where $u(z) = \frac{1}{2}[ z + \sqrt{z^2 - 4}]$.  So indeed
the result  (\ref{logmoment}) for $p= 1$ is consistent with the
previously known results (\ref{G2}) in the large N limit.
Note that the result (\ref{G2}) has been derived
by taking the large N limit first ; hence it is a smoothed correlation
function, which neglects all the  oscillatory  terms. (The connected
two-point correlation
function is $\rho_{2c}(\la_1,\la_2) = - \frac{1}{4\pi^2}[ G_{2c}
(\la_1+i\epsilon,\la_2+i\epsilon)
+G_{2c}(\la_1-i\epsilon,\la_2-i\epsilon) -
G_{2c}(\la_1+i\epsilon,\la_2-i\epsilon)
- G_{2c}(\la-i\epsilon,\la_2+i\epsilon)]$, and
it becomes $-1/2\pi^2 N^2 (\la_1 - \la_2)^2$ for $\la_1$ close to $\la_2$.
This result is obtained by  smoothing  the oscillatory part,
while the exact result is
$\rho_{2c}(\la_1,\la_2) \simeq - \sin^2 x/\pi^2 N^2 (\la_1 - \la_2)$; by
taking the large-N limit first, the $\sin^2x$
is replaced by $1/2$.).

We noticed that a similar formula exists for the Riemann $\zeta$-function,
although it deals with the imaginary part of the logarithm
of the $\zeta$-function,
from which a study of the variance of the number of zeros has been discussed
in the literature
\cite{Berry,Fujii}.


\begin{thebibliography}{99}

\bibitem{BH1} E. Br\'ezin and S. Hikami, a preprint,
math-ph/9910005.
\bibitem{Keating} J. Keating and N. Snaith, Lecture at Erwin Schrodinger
Institute,(1998).
\bibitem{Conrey} J. B. Conrey and S. M. Gonek, a preprint.
\bibitem{Heath-Brown} D. R. Heath-Brown,
Quart. J. Math. Oxford {\bf 44} (1991), 185-197.
\bibitem{Conrey-Farmer}
J. B. Conrey and D. W. Farmer,  a preprint.
\bibitem{BIPZ} E. Br\'ezin, C. Itzykson, G. Parisi and J.-B. Zuber
Comm. Math. Phys. {\bf 59} (1978), 35.
\bibitem{Selberg}
A. Selberg, Collected works Vol. I, P.353 and P.355, Springer, New York
(1989).
\bibitem{Bombieri} E. Bombieri and A. Peralli,
in {\it Analytic number theory}, edited by Y. Motohashi, London Mathematical
Society Lecture Note Series 247, Cambridge University Press (1997).
\bibitem{Laurincikas} A. Laurincikas, {\it Limit theorems for the
Riemann Zeta-function  }, Mathematics and Its applications Vol.352, Kluwer
Academic publishers, Dordrecht (1996).
\bibitem {Mehta} M.L. Mehta, {\it Random Matrices}, Academic Press,
New York (1991).
\bibitem{Dyson} F. J. Dyson, J. Math. Phys. {\bf 13}, 90 - 97 (1972).
\bibitem{BZ} E. Br\'ezin and A. Zee, Phys. Rev. E {\bf 49}, 2588 (1994).
\bibitem{BHZ} E. Br\'ezin, S. Hikami and A. Zee, Phys. Rev. E
{\bf 51}, 5442 (1995).
\bibitem{Berry} M. V. Berry, Nonlinearity {\bf 1}, 399-407 (1988).
\bibitem{Fujii}
A. Fujii, in {\it Emerging applications of number theory}, edited by D.
A. Hejhal et al. IMA 109, Springer-Verlag, New York (1999).
\end{thebibliography}
\end{document}